\documentclass[aps,prb,twocolumn,superscriptaddress,showpacs]{revtex4}
\usepackage{graphicx}
\usepackage{dcolumn}
\usepackage{bm}
\begin{document}
\title{On the origin of the non-monotonic doping dependence of the in-plane resistivity anisotropy of Ba(Fe$_{1-x}T_x$)$_2$As$_2$, $T$ = Co, Ni and Cu}
\def \lafepo{LaFePO}
\def \Ba122{BaFe$_2$As$_2$}
\def \CoBa122{Ba(Fe$_{1-x}$Co$_x$)$_2$As$_2$}
\def \NiBa122{Ba(Fe$_{1-x}$Ni$_x$)$_2$As$_2$}
\def \CuBa122{Ba(Fe$_{1-x}$Cu$_x$)$_2$As$_2$}
\def \CoEu122{Eu(Fe$_{1-x}$Co$_x$)$_2$As$_2$}
\def \LSCO{La$_{2-x}$Sr$_x$CuO$_4$}
\def \Tc{$T_c$}
\def \Ts{$T_s$}
\def \TN{$T_N$}
\author{Hsueh-Hui Kuo}
\thanks{Both authors contributed equally to this work.}
\affiliation{Department of Materials Science and Engineering and Geballe Laboratory for Advanced Materials, Stanford University, Stanford, California 94305, USA}
\affiliation{Stanford Institute of Energy and Materials Science, SLAC National Accelerator Laboratory, 2575 Sand Hill Road, Menlo Park 94025,California 94305, USA}
\author{Jiun-Haw Chu}
\thanks{Both authors contributed equally to this work.}
\affiliation{Stanford Institute of Energy and Materials Science, SLAC National Accelerator Laboratory, 2575 Sand Hill Road, Menlo Park 94025,California 94305, USA}
\affiliation{Department of Applied Physics and Geballe Laboratory for Advanced Materials, Stanford University, Stanford, California 94305, USA}
\author{Scott C. Riggs}
\affiliation{Stanford Institute of Energy and Materials Science, SLAC National Accelerator Laboratory, 2575 Sand Hill Road, Menlo Park 94025,California 94305, USA}
\affiliation{Department of Applied Physics and Geballe Laboratory for Advanced Materials, Stanford University, Stanford, California 94305, USA}
\author{Leo Yu}
\affiliation{E. L. Ginzton Laboratory, Stanford University, Stanford, California 94305, USA}
\author{Peter L. McMahon}
\affiliation{E. L. Ginzton Laboratory, Stanford University, Stanford, California 94305, USA}
\author{Kristiaan De Greve}
\affiliation{E. L. Ginzton Laboratory, Stanford University, Stanford, California 94305, USA}
\author{Yoshihisa Yamamoto}
\affiliation{E. L. Ginzton Laboratory, Stanford University, Stanford, California 94305, USA}
\affiliation{National Institute of Informatics, Hitotsubashi 2-1-2, Chiyoda-ku, Tokyo 101-8403, Japan}
\author{James G. Analytis}
\affiliation{Stanford Institute of Energy and Materials Science, SLAC National Accelerator Laboratory, 2575 Sand Hill Road, Menlo Park 94025,California 94305, USA}
\affiliation{Department of Applied Physics and Geballe Laboratory for Advanced Materials, Stanford University, Stanford, California 94305, USA}
\author{Ian R. Fisher}
\affiliation{Stanford Institute of Energy and Materials Science, SLAC National Accelerator Laboratory, 2575 Sand Hill Road, Menlo Park 94025,California 94305, USA}
\affiliation{Department of Applied Physics and Geballe Laboratory for Advanced Materials, Stanford University, Stanford, California 94305, USA}

\begin{abstract}
The in-plane resistivity anisotropy has been measured for detwinned single crystals of \NiBa122 and \CuBa122. The data reveal a non-monotonic doping dependence, similar to previous observations for \CoBa122. Magnetotransport measurements of the parent compound reveal a non-linear Hall coefficient and a large linear term in the transverse magnetoresistance. Both effects are rapidly suppressed with chemical substitution over a similar compositional range as the onset of the large in-plane resistivity anisotropy. This suggests that the relatively small in-plane anisotropy of the parent compound in the spin density wave state is due to the presence of an isotropic, high mobility pocket of the reconstructed Fermi surface. Progressive suppression of the contribution to the conductivity arising from this isotropic pocket with chemical substitution eventually reveals the underlying in-plane anisotropy associated with the remaining Fermi surface pockets. 
\end{abstract}


\maketitle

\section{Introduction}

Recent measurements of detwinned single crystals of \CoBa122 revealed a non-monotonic doping dependence of the in-plane resistivity anisotropy \cite{Chu_2010b}. In striking contrast, the lattice orthorhombicity diminishes monotonically with increasing Co concentration \cite{Prozorov_2009}, raising the question of the origin of the nonmonotonic behavior of the in-plane resistivity anisotropy, and the extent to which it is, or is not, generic to this family of compounds. Here we present measurements of the in-plane resistivity anisotropy of the closely related cases of Ni and Cu-substituted \Ba122 which reveal a similar non-monotonic compositional dependence of the resistivity anisotropy. Furthermore, magnetotransport measurements indicate that the effect is closely coupled to the progressive erosion of the contribution to the conductivity from an isotropic, high-mobility, reconstructed Fermi surface (FS) pocket.

\Ba122 is a representative ``parent" phase of the Fe-pnictide superconductors \cite{recent_reviews,Rotter_2008a,Rotter_2008b}. The material has an antiferromagnetic (AFM) ground state, comprising stripes of ferromagnetically aligned moments which alternate antiferromagnetically along the orthorhombic $a$-axis \cite{Huang_2008}. Back folding of the bands according to the antiferromagnetic wavevector results in a reconstructed Fermi surface consisting of several small pockets, as evidenced by both Angle Resolved PhotoEmission Spectroscopy (ARPES) \cite{Yi_2009a} and quantum oscillations \cite{Analytis_2009,Terashima_2011}. The material has a Neel temperature \TN\ close to 140 K, with values depending slightly on growth conditions and annealing treatments \cite{Rotter_2008a,Sefat_2008,Ni_2008,Chu_2009, Rotundu_2010}. Significantly, the Neel transition in \Ba122 is accompanied by a tetragonal-to-orthorhombic structural transition \cite{Huang_2008}. For the specific cases of Co, Ni and Cu substitution relevant to the current work, the structural transition occurs at a slightly higher temperature \Ts\ than the magnetic transition, with a temperature difference that monotonically increases with increasing concentration of the substituent, at least until the top of the superconducting dome \cite{Ni_2008,Chu_2009,Canfield_2009,Canfield_2010,Lester_2009,Ni_2010}. The origin of the splitting of \Ts\ and \TN\ with chemical substitution is not clear, but consideration of the effect of crystal quality on the splitting of the transitions in CeFeAsO \cite{Jesche_2010} implies that this effect might, at least in part, be associated with the strong in-plane disorder introduced by partial substitution on the Fe site. In both families, the structural transition breaks a discrete rotational symmetry ($C_4$ to $C_2$) of the high-temperature phase without introducing a new translational symmetry, and is widely referred to as a nematic transition, borrowing language from the field of liquid crystals\cite{Fradkin_2010}. Understanding the origin of this effect is a key component of a complete theoretical description of the occurrence of superconductivity in this family of compounds, motivating both theoretical\cite{Theory,Theory2,Chen_2010} and experimental\cite{Tanatar_2010,Blomberg_2011,Ying_2010,Yi_2010,Kim_2010,Dusza_2010,Exp} investigation of the nematic transition and the associated in-plane anisotropy.

\Ba122 tends to form dense structural twins on cooling through \Ts, corresponding to alternation of the orthorhombic $a$ and $b$ axes through the crystal \cite{Tanatar_2009}. The relative twin population can be influenced by application of an in-plane magnetic field due to the in-plane susceptibility anisotropy associated with the colinear antiferromagnetic structure \cite{Chu_2010a}. However, the degree of detwinning that can be achieved for typical laboratory fields is only modest and the anisotropy can only be explored for temperatures below $T_N$ \cite{Chu_2010a}. Much larger changes in the relative twin domain population can be achieved by use of uniaxial mechanical stress, which also permits measurement of the resistivity anisotropy through $T_s$ \cite{Chu_2010b,Tanatar_2010,Ying_2010,Blomberg_2011}. Such measurements reveal a relatively small in-plane resistivity anisotropy in the parent compound \Ba122 \cite{Chu_2010b,Tanatar_2010,Blomberg_2011}, with the resistivity along the ferromagnetic direction $\rho_{b}$ slightly greater than that along the antiferromagnetic direction $\rho_{a}$. In contrast, the anisotropy $\rho_b/\rho_a$ is initially found to increase with Co substitution, despite the fact that the orthorhombicity $(a-b)/2(a+b)$ monotonically decreases with increasing Co concentration \cite{Chu_2010b,Ying_2010}. As anticipated, the anisotropy eventually diminishes to unity when $T_s$ is completely suppressed. Perhaps coincidentally, the maximum in $\rho_b/\rho_a$ is found to occur for a Co concentration close to the beginning of the superconducting ``dome" \cite{Chu_2010b}.

Optical reflectivity measurements for mechanically detwinned crystals reveal that the origin of the resistivity anisotropy is principally caused by changes in the spectral weight (i.e. is due to changes in the FS morphology), rather than by changes in the scattering \cite{Dusza_2010}. The anisotropy is present over a wide energy scale, clearly involving whole bands rather than just the behavior at the Fermi energy, and the dichroism is largest for the parent compound, consistent with the structural orthorhombicity \cite{Dusza_2010}. In addition, recent polarized ARPES measurements on detwinned crystals reveal that the structural transition is associated with an increase (decrease) in the binding energy of bands with principal $d_{xz}$ ($d_{yz}$) character \cite{Yi_2010, Kim_2010}. The relative degree of splitting of the two bands is largest for the parent compound and decreases monotonically with increasing Co concentration.

Despite these recent advances, the origin and significance of the nonmonotonic doping dependence of the in-plane resistivity anisotropy in \CoBa122 remains unclear. Initial calculations with a net orbital polarization have suggested the magnitude of resistivity anisotropy could strongly depend on the density of states near the Fermi level, which does not necessarily have a monotonic doping dependence\cite{Chen_2010}. In this regard, it is especially useful to consider other dopants in order to establish systematic trends. In this paper, we compare the temperature and doping dependence of the in-plane resistivity anisotropy of detwinned crystals of Co, Ni and Cu substituted \Ba122. All three reveal a non-monotonic doping dependence to $\rho_b$/$\rho_a$, but with the maximum in-plane anisotropy occurring for different ranges of the dopant concentration. For the two cases of Co and Ni substitution, for which a direct comparison of the physical properties is best motivated both phenomenologically and also based on ab-initio calculations \cite{Arita_2010}, the onset of the large in-plane resistivity anisotropy coincides with a suppression of both a large \textit{non-linear} contribution to the transverse resistivity $\rho_{xy}$, and also of a \textit{linear} term in the magnetoresistance, suggesting a common origin for the two effects.

\section{Experimental Methods}

\begin{figure}
\includegraphics[width=8.5cm]{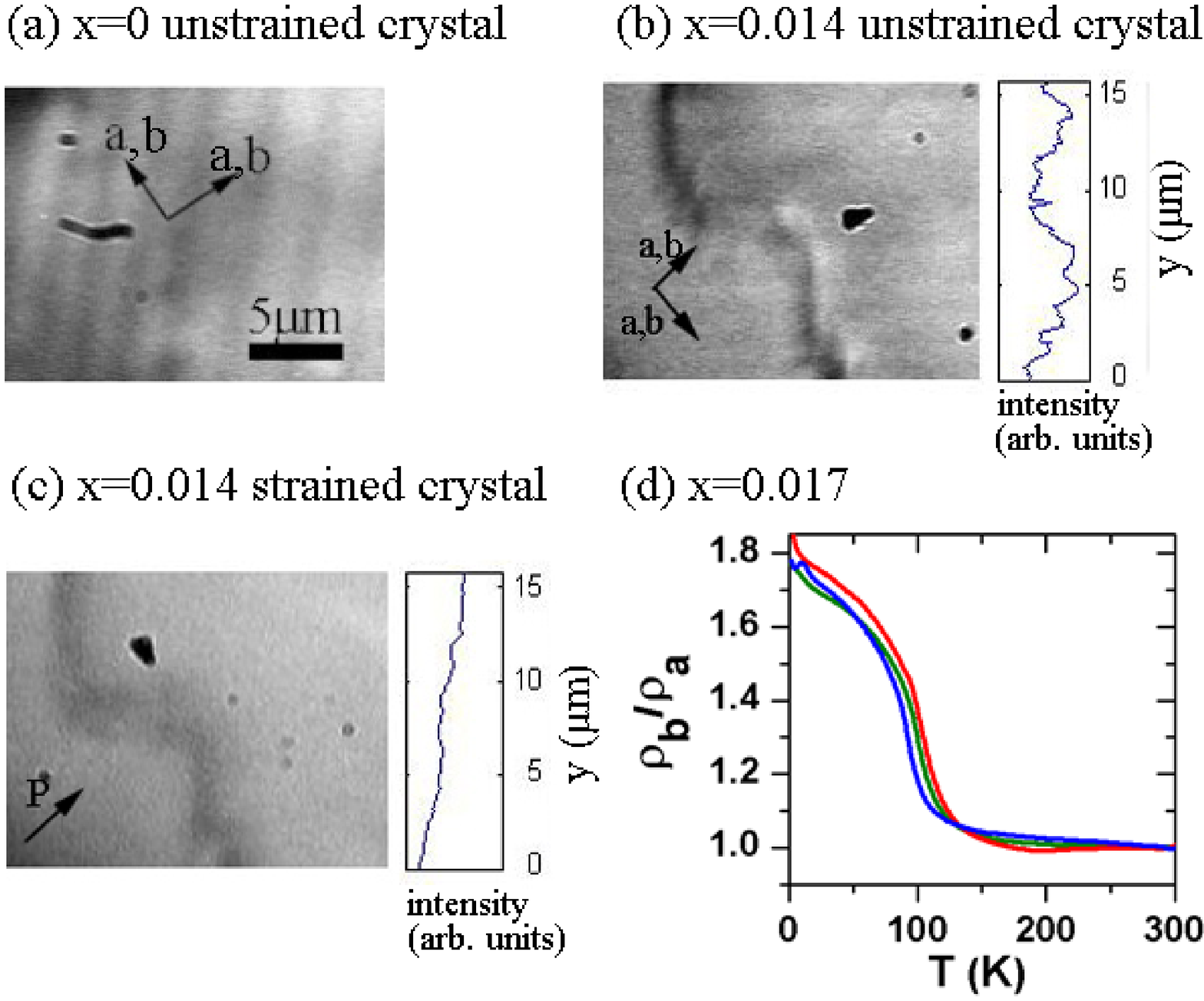}
\caption{\label{Opt} (Color online) (a) Image of the surface of an unstressed single crystal of \Ba122 at a temperature below 10 K, obtained using polarized-light microscopy, as described in the main text. Stripes of light and dark contrast are associated with different twin orientations. Arrows indicate the orientation of crystal $a$ and $b$ axes. (b) Image of an unstressed crystal of \NiBa122 with $x$=0.014, also revealing horizontal stripes corresponding to the two twin orientations, but with a significantly reduced contrast. The jagged feature running from top to bottom of the image is due to surface morphology. To better reveal the stripes, the intensity was integrated in the horizontal direction and plotted as a function of vertical position, $y$ (right hand axis). (c) The same region of the same crystal as shown in panel (b), but with uniaxial stress applied in the direction indicated, revealing the detwinning effect of the uniaxial stress. Note the absence of horizontal stripes in both the photograph and also the integrated intensity plot. (d) The in-plane resistivity anisotropy expressed as $\rho_b/\rho_a$ for three different crystals of \NiBa122 with $x \sim 0.017$, illustrating that the  measurements are reproducible.}
\end{figure}

Single crystals of \NiBa122 and \CuBa122 were grown from a self-flux \cite{Chu_2009, Sefat_2008}. Ba was combined with a mixture of FeAs and Ni/Cu with a ratio of Ba:FeAs:Ni/Cu = 1:4:x. The mixture was held in an alumina crucible and sealed in quartz, and was heated slowly to 1190 $^\circ$C and then cooled down to 1000 $^\circ$C in 60 hrs, at which temperature the remaining flux was decanted. The Ni and Cu content of the resulting crystals was measured by electron microprobe analysis (EMPA) using \Ba122 and elemental Ni and Cu as standards. Measurements were made for several points on each crystal, with standard deviations for the Ni and Cu concentrations which were generally below 10\% and 14\% of the absolute values respectively.

A mechanical detwinning device, with a design similar to that which was previously described for our earlier measurements of \CoBa122 \cite{Chu_2010b} but modified to be suitable for somewhat smaller crystals of \NiBa122 and \CuBa122, was used to mount the crystals for measurement of the in-plane resistivity. Crystals were cut into rectilinear bars with the tetragonal $a$-axis at 45 degrees to the sides of the bar (corresponding to the natural crystal facets), and with in-plane aspect ratios of approximately 1:1.2. The crystals were placed on a horizontal platform on the detwinning device, such that an insulated Cu plate rested on the edge of the crystal. Uniaxial pressure was applied by tightening the Cu plate against the edge of the crystal. A similar magnitude stress was applied to all crystals studied, but absolute values could not be estimated for the small cantilevers used in this study. On cooling through $T_s$, the uniaxial pressure favors the twin orientation with the shorter $b$-axis along the direction of the applied compressive stress \cite{Chu_2010b}.

The degree of detwinning was monitored for several representative crystals via polarized light microscopy. Samples were positioned on the cold finger of a vacuum cryostat. The samples were illuminated with $\sim$800 nm light, linearly polarized at approximately 45 degrees to the orthorhombic a/b-axes. The reflected light was passed through an optical compensator and analyzed by an almost fully crossed polarizer to maximize the contrast in birefringence between the two twin orientations. Whereas good contrast between the two twin orientations is possible for the parent compound (Fig.\ref{Opt}(a)) and for \CoBa122 for compositions across the phase diagram \cite{Chu_2010a}, Ni and Cu substitution are found to rapidly suppress the contrast between the two twin orientations. Representative images for a single crystal of \NiBa122 with $x$ = 0.014 with and without uniaxial stress are shown in Fig.\ref{Opt} (b) and (c) respectively. The application of uniaxial stress clearly results in a single twin domain orientation over the field of view, in this case about 22 microns. The origin of the reduced contrast relative to \CoBa122 is not known. The twin domain dimensions seen in Figs.\ref{Opt}(b) are similar to those found for \CoBa122, so it is unlikely that the width of the domains for higher Ni concentrations fall below the resolution of the microscope, although we cannot completely exclude this possibility. Although the optical imaging was not possible for $x$ $>$ 0.014 for Ni-substituted samples, or for $x$ $>$ 0 for Cu substituted samples, a clear change in the resistivity was observed as a consequence of the applied stress for all underdoped compositions. Since we are unable to independently determine the degree of detwinning for these compositions, the observed in-plane resistivity anisotropy must therefore be considered a minimum bound on the actual value. However, for several representative compositions, measurements were made with multiple crystals, in each case revealing almost identical changes in the in-plane resistivity for stress applied parallel and perpendicular to the current (Fig.\ref{Opt}(d) for example), providing evidence that even for the samples for which optical characterization failing to verify the effect of detwinning, the samples were essentially fully detwinned.

The in-plane resistivity was measured using a standard 4-probe configuration. Crystals were rotated so that the applied pressure was parallel and perpendicular to the current, enabling measurement of $\rho_{b}$ and $\rho_{a}$ respectively. The same contacts were used for each measurement, to avoid uncertainty in the geometric factors. Measurements were made for multiple crystals of several compositions to ensure reproducibility of the results.  The in-plane resistivity was also measured for unstrained crystals in order to determine $T_s$ for each composition. For several samples, the in-plane resistivity after straining was also measured, and no change in either \Ts\ or \TN\ was observed.
Finally, both $\rho_{xx}$ and the transverse resistivity $\rho_{xy}$ were measured at the National High Magnetic Field Laboratory (NHMFL) in Tallahassee in dc magnetic fields up to 35 T and in Stanford in fields up to 14 T. The magnetic field was always oriented along the $c$-axis and the samples were mounted using a 6-point contact configuration. Measurements were made for both positive and negative field orientations in order to subtract any small resistive component due to contact misalignment. 

\section{Results}

\begin{figure*}[th]
\includegraphics[width=8.5cm]{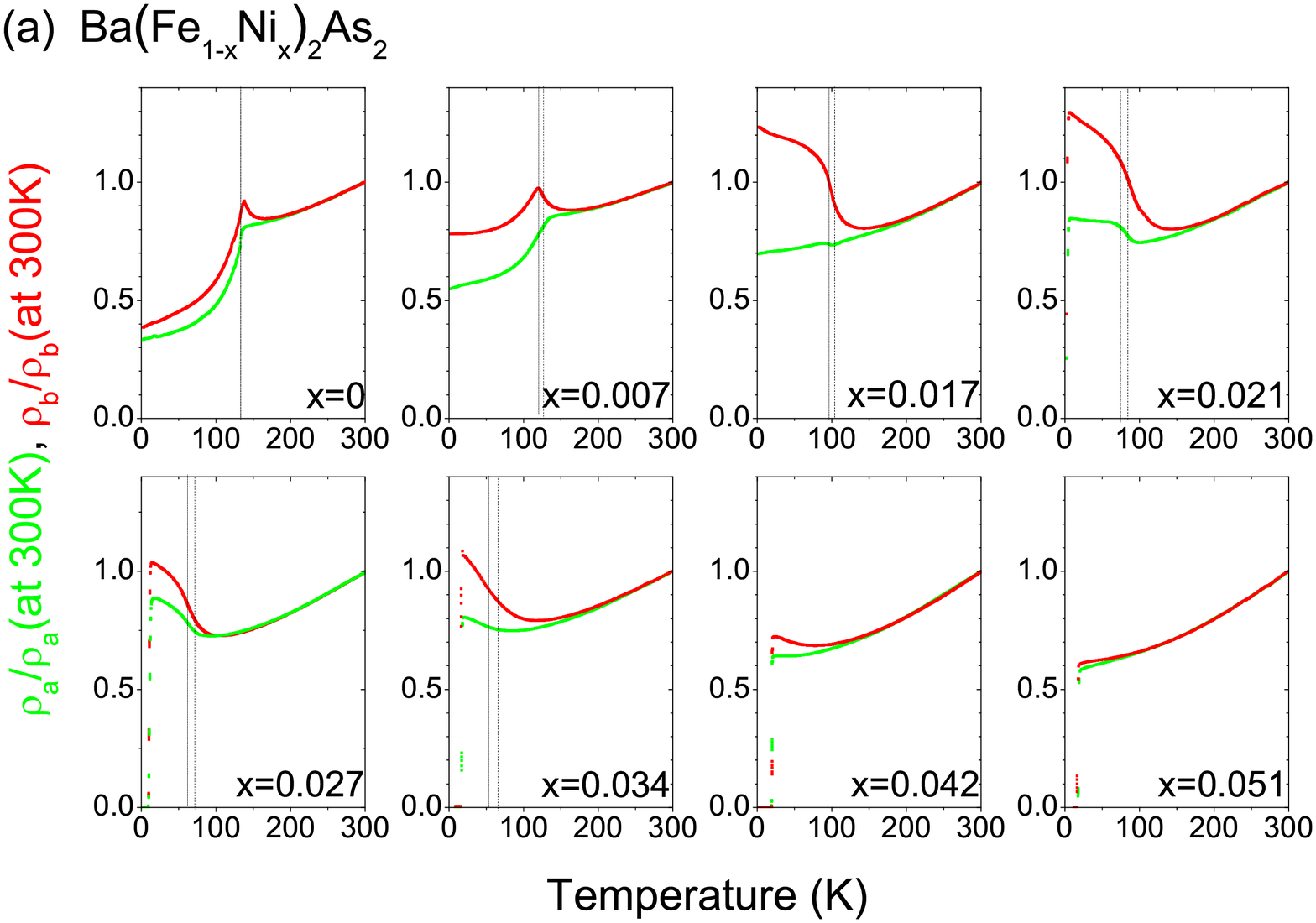}
\includegraphics[width=8.5cm]{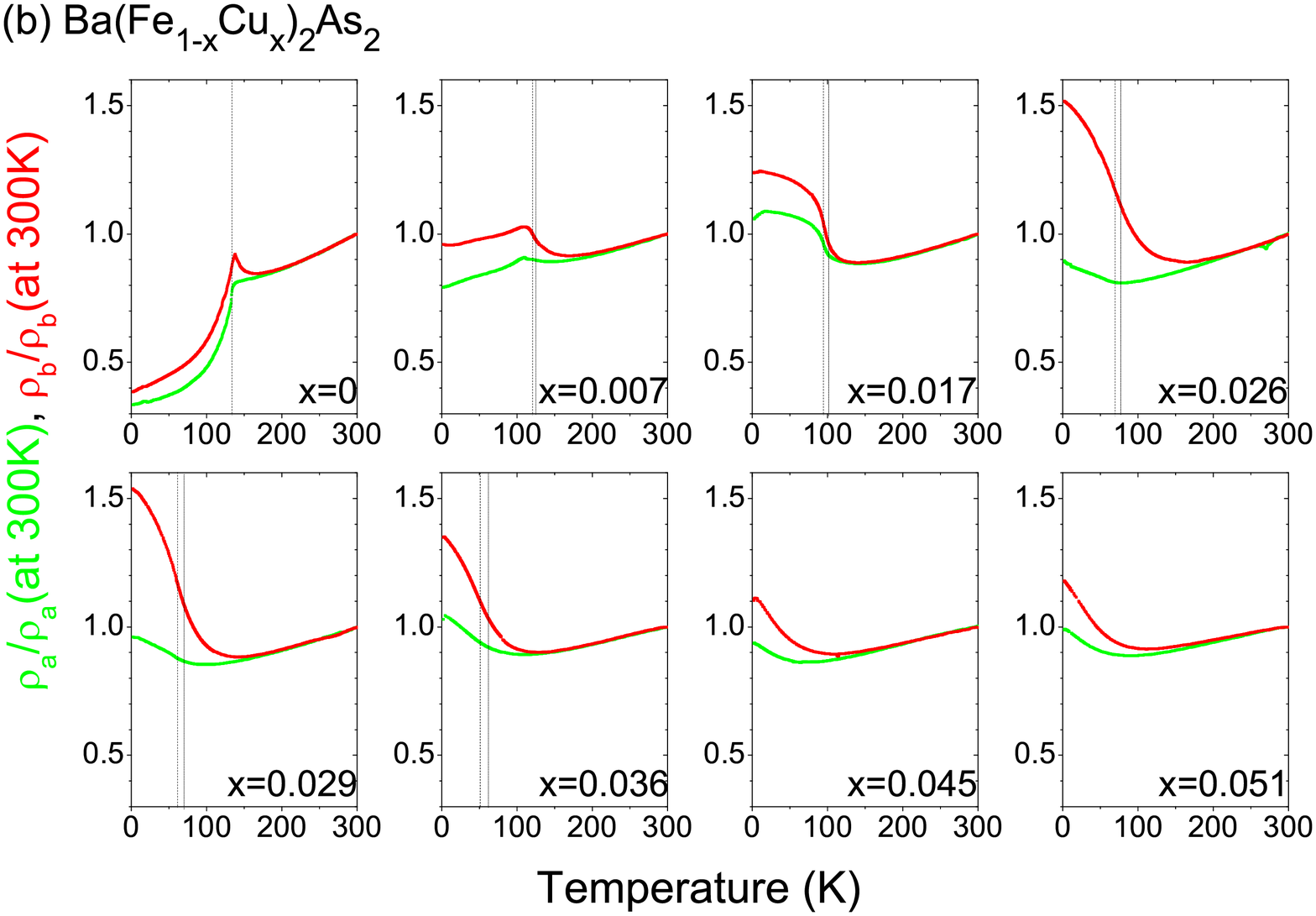}
\caption{\label{Rba} (Color online) Temperature dependence of the in-plane resistivity $\rho_{a}$ (green curves) and $\rho_{b}$ (red curves) of stressed crystals of (a) \NiBa122  and (b) \CuBa122. Data have been normalized by the value at 300 K. Values of $x$ are label-led in each panel. Vertical lines mark \Ts (dot-dashed line) and \TN (dashed line) determined from $d\rho(T)/dT$ for unstressed conditions.} 
\end{figure*}

\begin{figure}
\includegraphics[width=8.5cm]{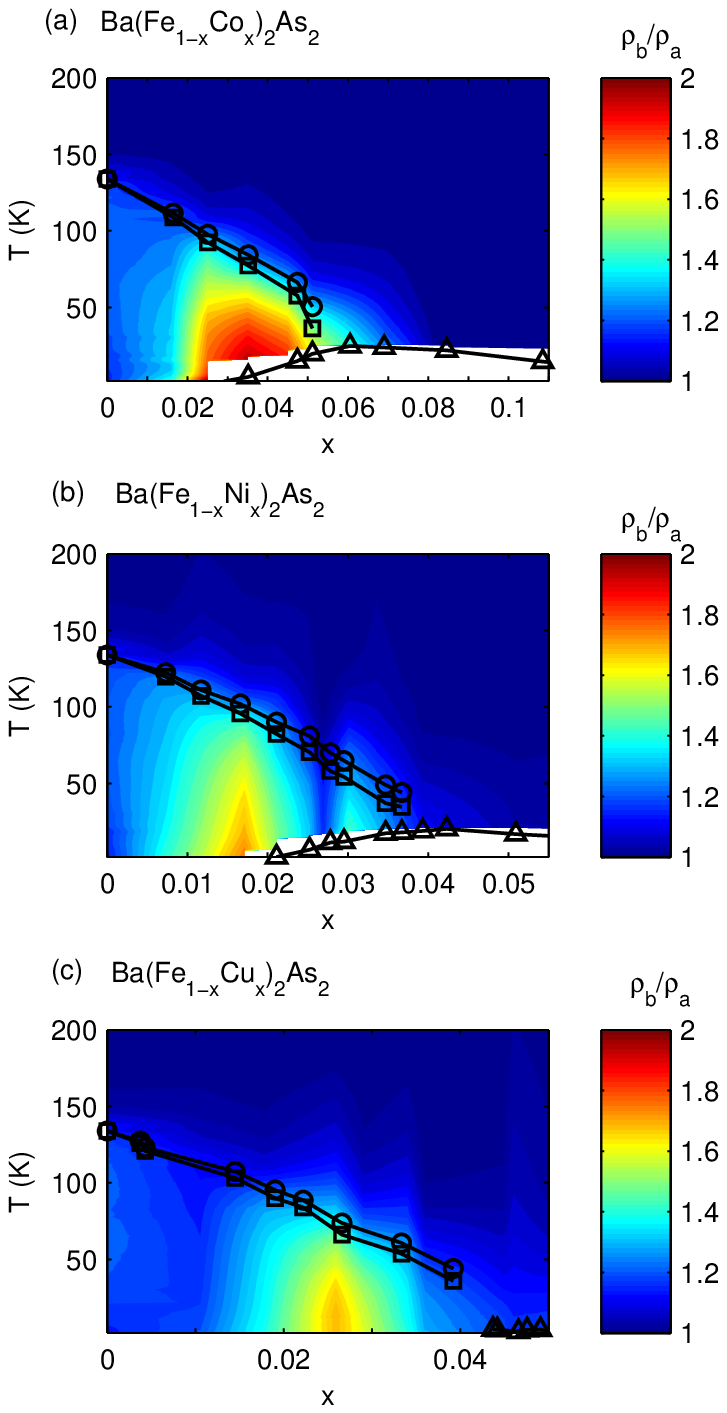}
\caption{\label{CoNiCu} (Color online) In-plane resistivity anisotropy, $\rho_{b}$/$\rho_{a}$, as a function of temperature and doping for (a) \CoBa122, (b) \NiBa122, and (c) \CuBa122. The color scale has been obtained by a linear interpolation between adjacent data points. The same scale has been used for all three panels. Black circles, Squares, and triangles indicate \Ts , \TN , and \Tc\ respectively, determined for unstressed conditions. \Tc\ for \CoBa122 and \NiBa122 was defined by the midpoint of the superconducting transitions, while it represents the onset temperature of the transition for \CuBa122.}
\end{figure}

Results of measurements of the in-plane resistivity for representative single crystals of \NiBa122 and \CuBa122 held under an applied uniaxial stress in the detwinning device are shown in Figs.\ref{Rba}(a) and (b) respectively. For each composition, $\rho_{a}$ and $\rho_{b}$ were measured for the same crystal. Nevertheless, the small crystals used for these measurements are susceptible to damage while being repositioned in the detwinning device, so the resistivity has been normalized by its value at 300 K in order to avoid uncertainty arising from subtle changes in the geometric factors between each measurement. For cases for which the applied stress is perpendicular to the current, the data are labeled as $\rho_a$ (green curves), whereas for cases for which the applied stress is parallel to the current the data are label-led as $\rho_b$ (red curves). As described previously \cite{Chu_2010b}, $T_N$ is unaffected by the small stress used to detwin the crystals, but the structural transition is rapidly broadened. Vertical lines in Fig.\ref{Rba} mark $T_N$ and $T_s$ under conditions of zero stress.

As found previously for the undoped parent compounds and for \CoBa122 \cite{Chu_2010a,Chu_2010b,Tanatar_2010,Ying_2010,Blomberg_2011}, $\rho_b > \rho_a$ for all underdoped compositions. The difference begins gradually at a temperature well above $T_s$, but there is no indication in either the resistivity or its derivatives of an additional phase transition marking the onset of this behavior. The temperature at which the difference becomes discernible depends on pressure \cite{Chu_2010b}, and the effect appears to be associated with a large Ising nematic susceptibility. For both series, $\rho_b$ rapidly develops a steep upturn with decreasing temperature as the dopant concentration is increased from zero, similar to the behavior observed for \CoBa122. However, although $\rho_{a}$ in \CoBa122 exhibits metallic behavior up to $x = 0.035$, $\rho_a$ starts increasing with decreasing temperature in both \NiBa122 (beginning at $x \sim 0.018$) and in \CuBa122 (beginning at $x \sim 0.012$). This behavior is unlikely to be associated with partial mixing of the $b$-axis resistivity (for instance due to incomplete detwinning) because $\rho_a$ is found to increase with decreasing temperature even above $T_s$ for higher Ni and Cu concentrations (i.e. at a temperature for which there is no twin formation). 

\begin{figure}[t]
\includegraphics[width=8.5cm]{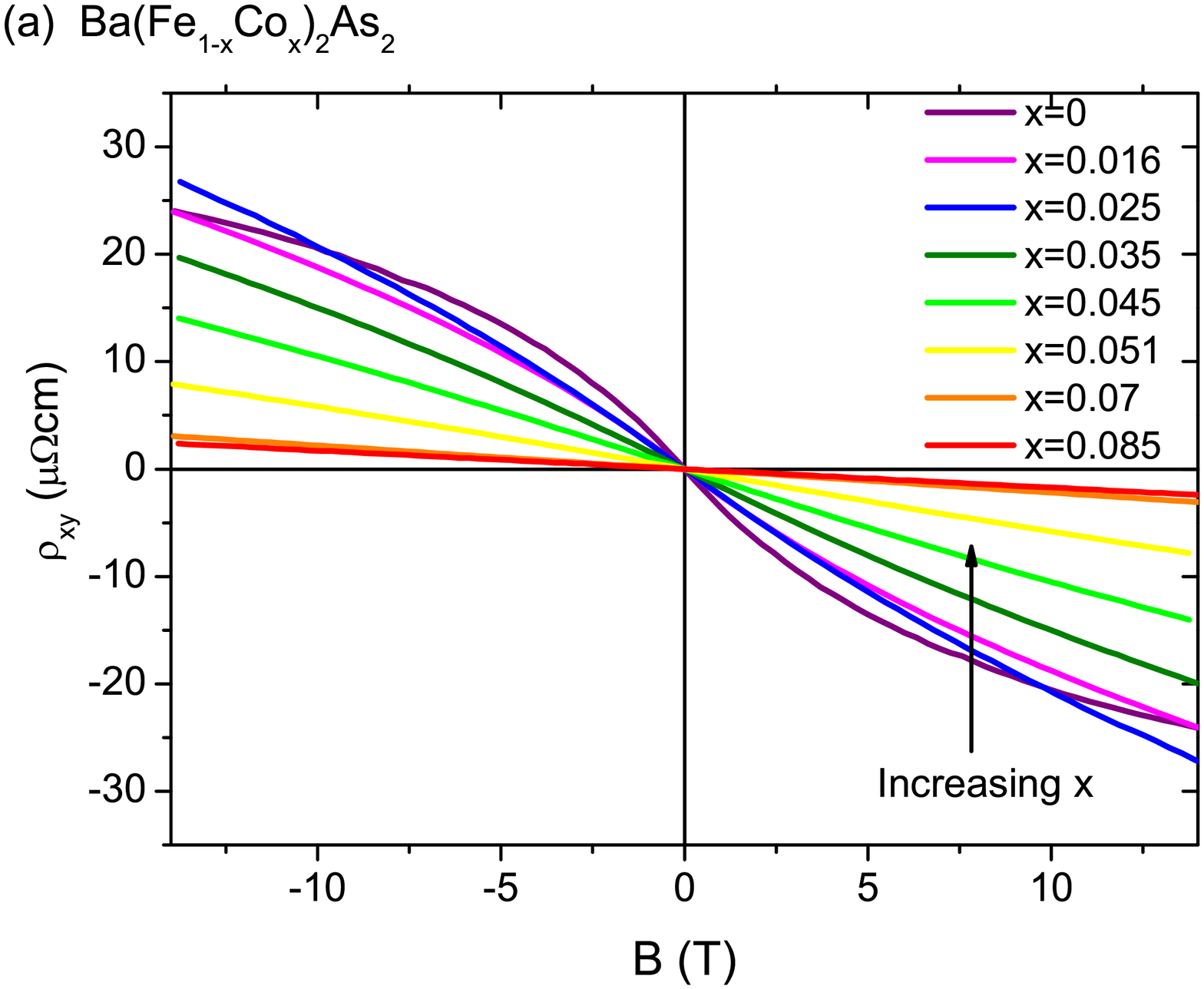}
\includegraphics[width=8.5cm]{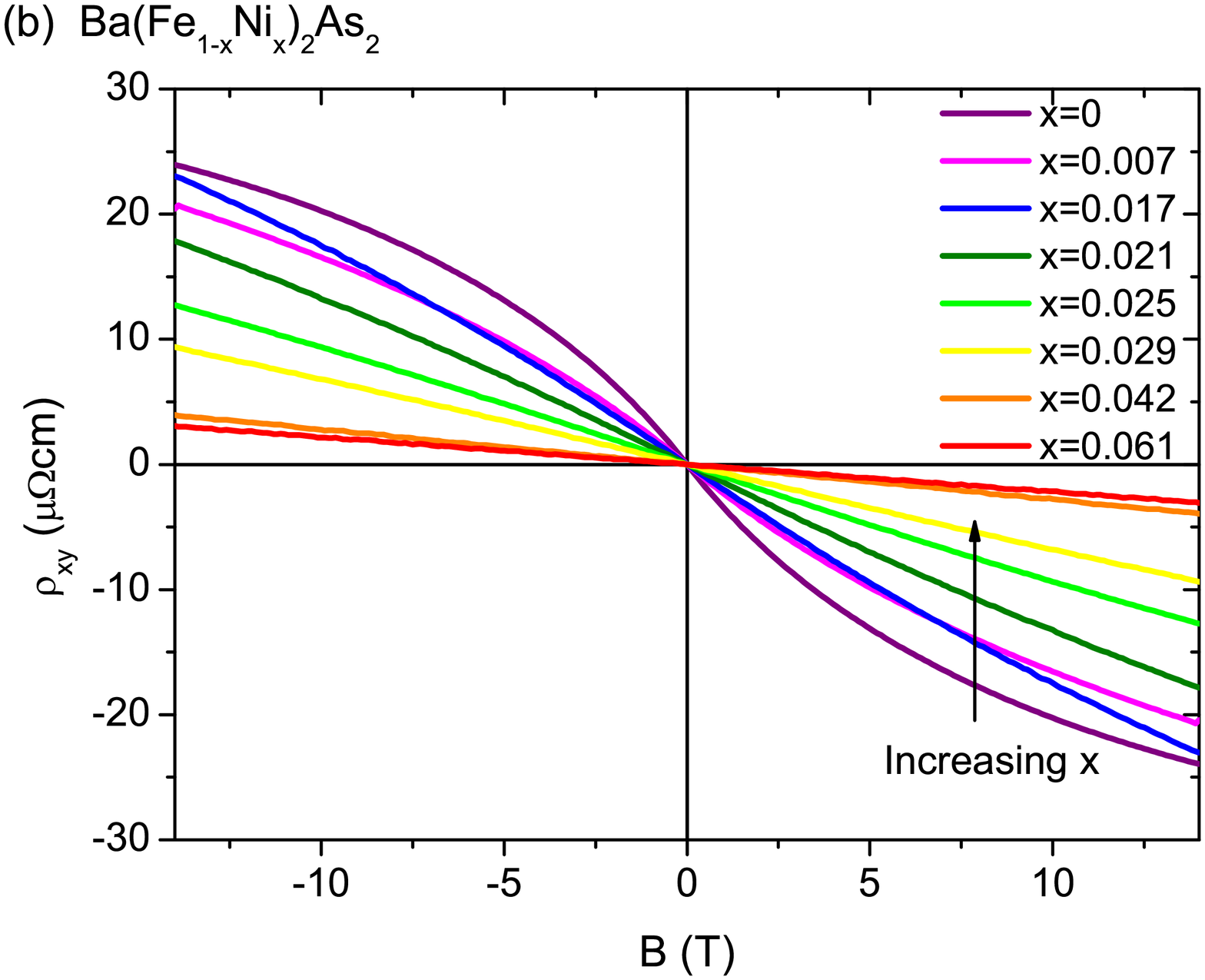}
\includegraphics[width=8.5cm]{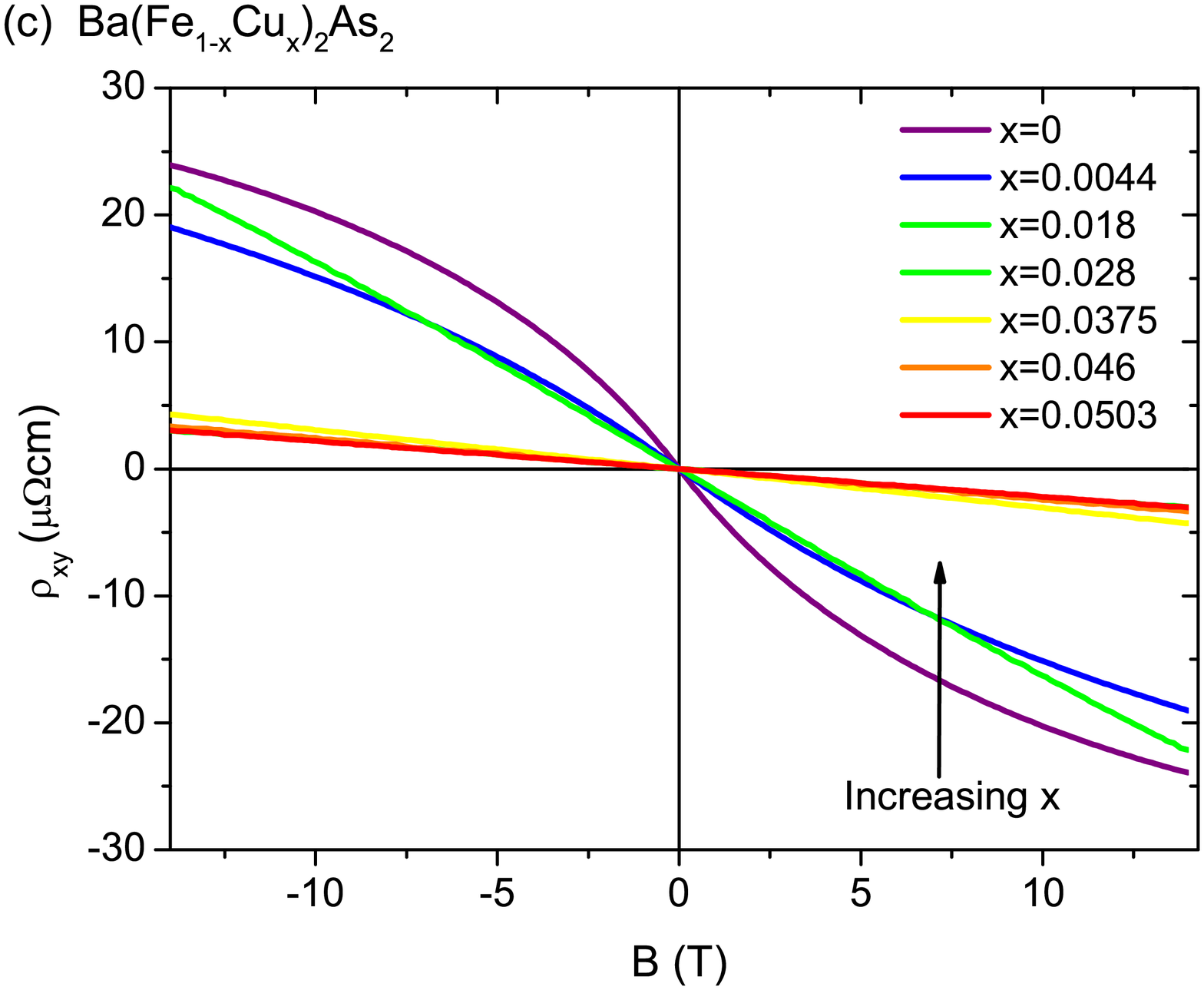}
\caption{\label{RxyH} (Color online) Field dependence of the transverse resistivity $\rho_{xy}$ at a temperature $T$ = 25$K$ with $B$ $\|$ $c$ for (a) \CoBa122, (b) \NiBa122 and (c) \CuBa122. Compositions are labeled in the legends.}
\end{figure}

The in-plane resistivity anisotropy, expressed as $\rho_b/\rho_a$, is shown as a function of temperature and composition in Fig.\ref{CoNiCu}. A linear interpolation between data points has been used to generate the color scale images. Data points indicate values of \Ts , \TN\ and \Tc\ under conditions of zero applied stress. For comparison, data for \CoBa122 taken from ref. \onlinecite{Chu_2010b}, are also shown. In all three cases, $\rho_b/\rho_a$ is found to vary non-monotonically with increasing amounts of Co, Ni or Cu. For the case of Co substitution, the in-plane anisotropy peaks at a value of nearly 2 for a composition $0.025 < x < 0.045$, close to the onset of the superconducting dome. Uncertainty in the exact composition at which $\rho_b/\rho_a$ is maximal reflects the relatively sparse data density. In comparison, for Ni substitution, $\rho_b/\rho_a$ peaks for 0.012 $<$ $x$ $<$ 0.022, approximately half the dopant concentration as for Co substitution. In addition, a much weaker secondary maximum is found for \NiBa122 centered at $x \sim$ 0.03. Measurements of multiple crystals confirmed the presence of this feature. For Cu substitution, the in-plane anisotropy peaks in the range 0.022 $<$ $x$ $<$ 0.03.

To investigate the origin of the non-monotonic doping dependence of the in-plane resistivity anisotropy, which is observed for all three dopants, we turn now to the results of magnetotransport measurements, starting with the Hall effect (Fig.\ref{RxyH}). For the parent compound \Ba122 , it has been well established that even in modest magnetic fields the transverse resistivity, $\rho_{xy}$ is non-linear\cite{Cooper_2009}. Non-linearity in $\rho_{xy}$ is expected in multi-band systems in which at least one FS pocket is not in the weak field limit.

\begin{figure}[ht]
\includegraphics[width=8.5cm]{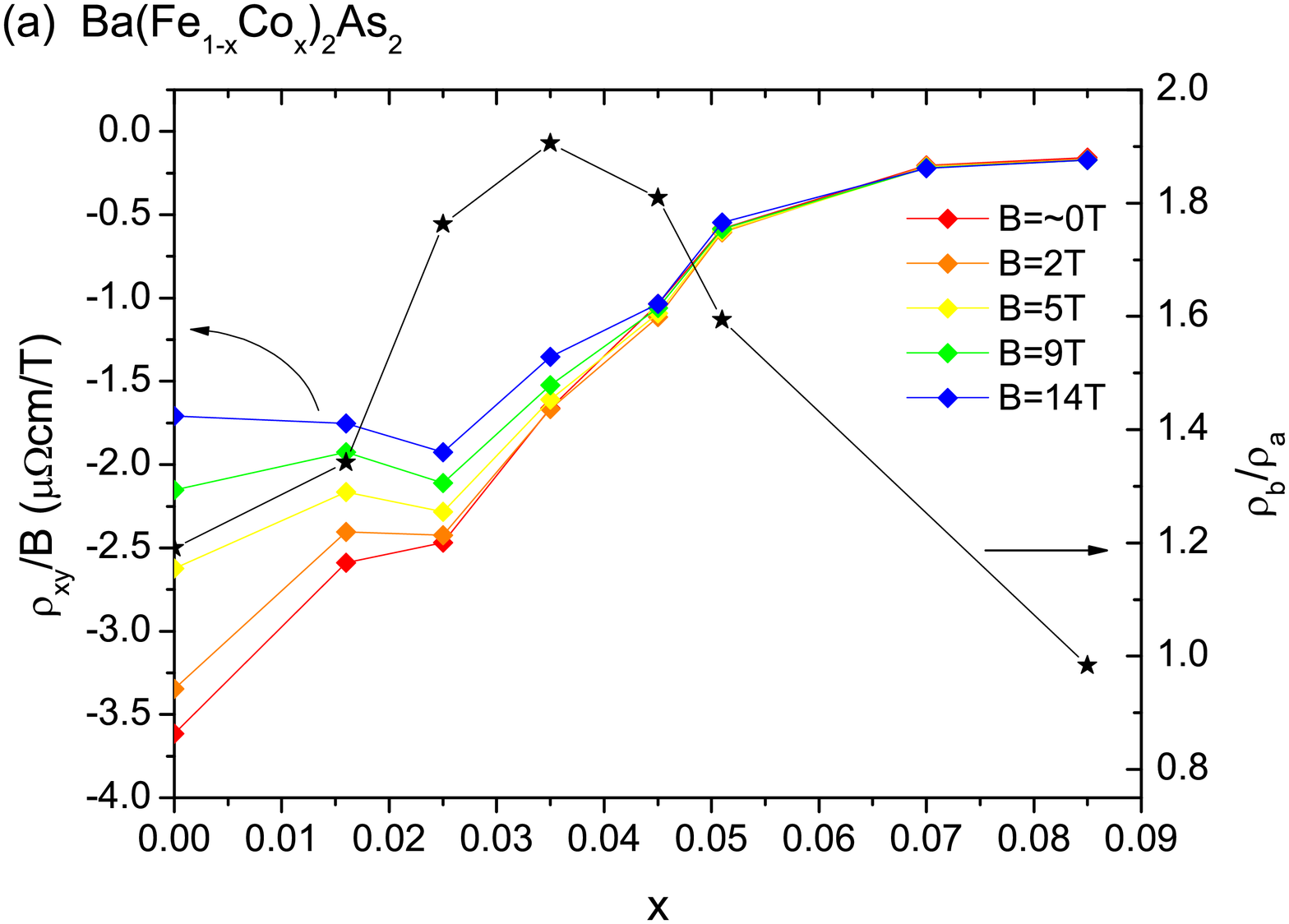}
\includegraphics[width=8.5cm]{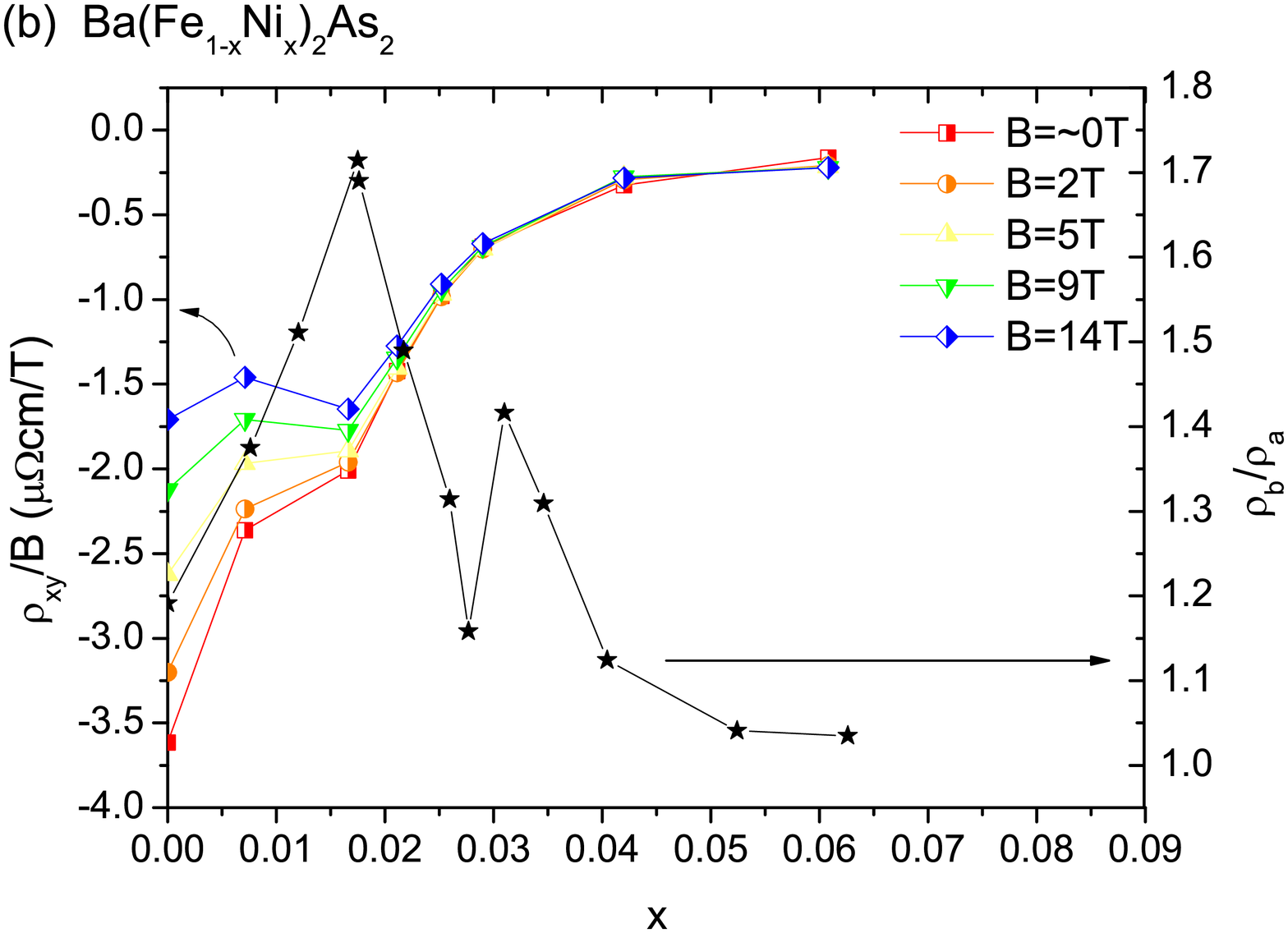}
\includegraphics[width=8.5cm]{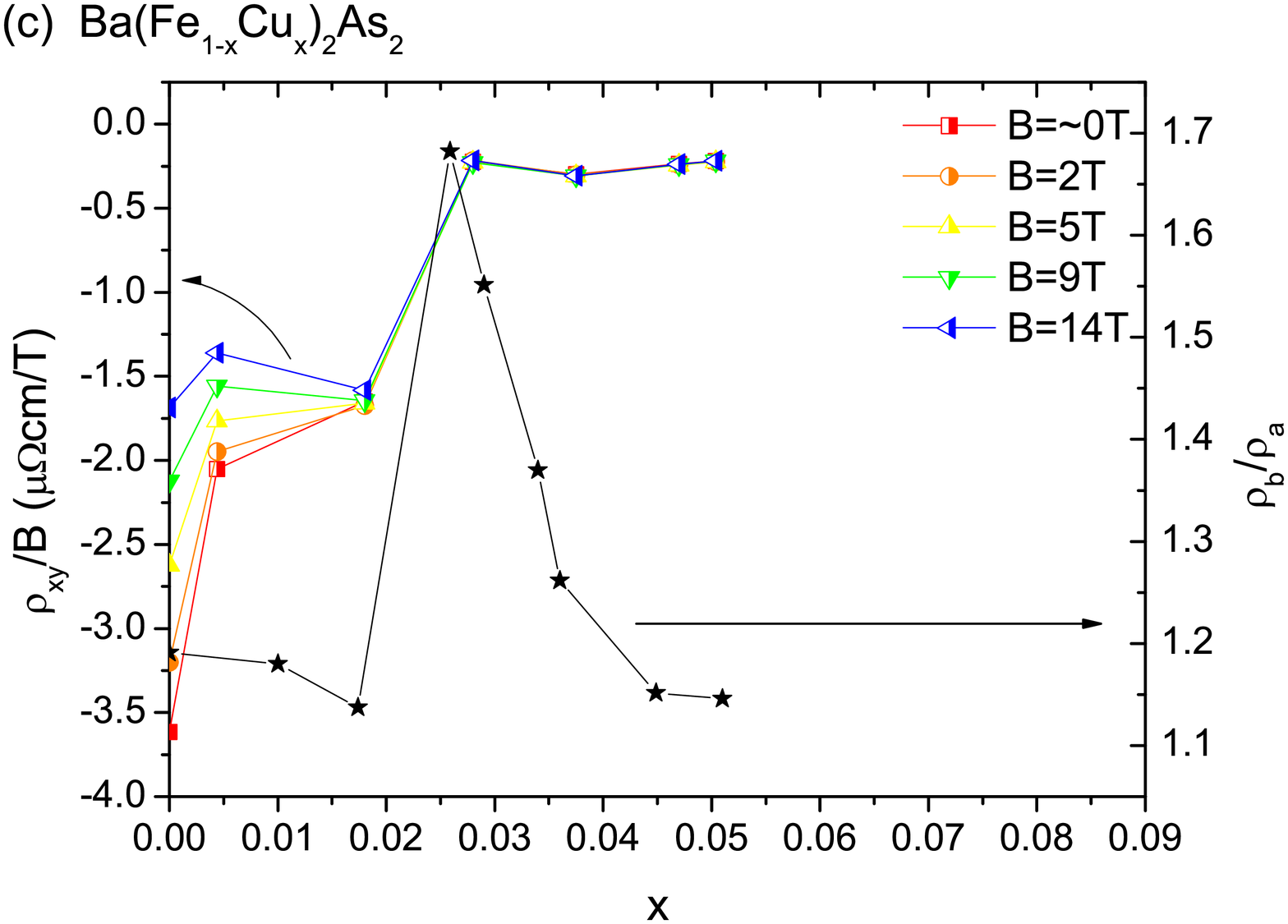}
\caption{\label{Rxy} (Color online) Left axes: composition-dependence of $\rho_{xy}$/$B$ at 25 K for (a) \CoBa122, (b) \NiBa122, and (C) \CuBa122. Values of $\rho_{xy}$/$B$ were evaluated at 0, 2, 5, 9 and 14 T, as described in the main text. Right axes (black data points): composition dependence of $\rho_b/\rho_a$ at 25 K.}
\end{figure}

Addition of either Co, Ni or Cu (Fig.\ref{RxyH}(a), (b) and (c) respectively) rapidly suppresses the non-linear behavior of $\rho_{xy}$. The rate at which the non-linear field dependence of $\rho_{xy}$ is suppressed with $x$ is best seen by considering $\rho_{xy}/B$ as a function of $x$ for different values of $B$. If $\rho_{xy}$ varies linearly with $B$, $\rho_{xy}/B$ yields a constant value, which is just the Hall coefficient, $R_H$. Data for all three series are shown in Fig.\ref{Rxy} as a function of $x$ for $T$ = 25 K, where $\rho_{xy}/B$ has been evaluated for representative fields $B$ = 0, 2, 5, 9 and 14 T. (Data for $B$ = 0 were evaluated by considering the instantaneous slope of $\rho_{xy}$ at $B$ = 0.) For the available field range (0$\sim$14 T), $\rho_{xy}/B$ becomes independent of field for $x >$ 0.045, 0.025 and 0.018 for Co, Ni and Cu substitution respectively. Clearly Cu substitution is more effective at suppressing the non-linear Hall behavior than Ni substitution, and Ni substitution is more effective than Co substitution. 

It is instructive to compare the composition dependence of $\rho_{xy}/B$ with that of the in-plane resistivity anisotropy, $\rho_b/\rho_a$. Black data points in Fig.\ref{Rxy} show $\rho_b/\rho_a$ (referenced to the right-hand axis) also evaluated at 25 K as a function of $x$ for all three substitution series. Comparison of data for Co and Ni substitution reveal that suppression of the non-linear behavior of $\rho_{xy}$ appears to be correlated with the onset of the large in-plane resistivity anisotropy. That is, Ni substitution seems to be almost twice as effective at both suppressing the non-linear Hall effect, and also at yielding a large in-plane anisotropy, than is Co substitution. This apparent correlation is less pronounced for the case of Cu substitution, for which, despite an even more rapid suppression of the non-linear Hall effect, the resistivity anisotropy appears to peak at a comparable range of compositions as found for Ni substitution.

\begin{figure}
\includegraphics[width=8.5cm]{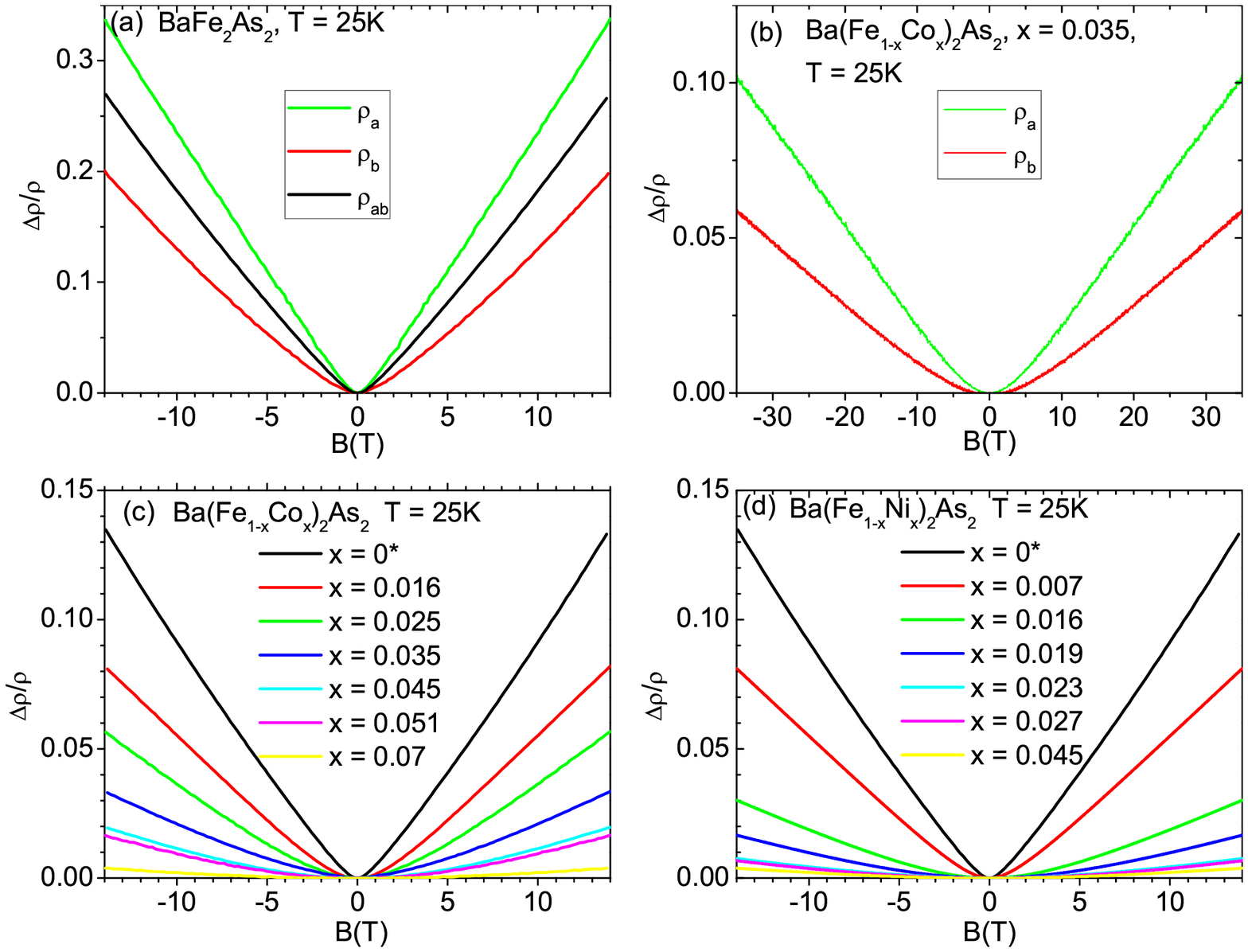}
\caption{\label{Fig:MR} (Color online) Transverse magnetoresistance($MR$) ($\Delta\rho/\rho$) of \CoBa122 for (a) $x$ = 0 (b) 3.5\% at $T = 25\ K$. The $MR$ has been measured on detwinned samples with current along $a$/$b$ axis (Green/Red curves), and also on twinned sample(Black curve). (c,d) The $MR$ of twinned samples of \CoBa122 and \NiBa122 respectively at 25 K. *Data for $x$ = 0 has been scaled down by a factor of two for clarity.}
\end{figure}

To shed more light on the correlation of the non-linear Hall effect and in-plane resistivity anisotropy, the transverse ($B$ parallel to $c$-axis) magnetoresistance (MR, defined as $\Delta\rho$/$\rho$) has also been measured on both detwinned and twinned samples. We first discuss the parent compound, for which representative MR data are shown in Fig.\ref{Fig:MR}(a) for a temperature of 25 K. As has been previously observed, the MR of the parent compound is linear over a wide field range \cite{Huynh_2010}. This behavior extends to very low fields, at which point the 	MR naturally reduces to a weak-field quadratic dependence. Interestingly, the linear behavior does not depend strongly on the current direction, and the difference of the linear slope can be mainly accounted for by the difference of the zero field resistivity, i.e. $\Delta\rho/\rho$ scales approximately with $B/\rho$. As was first shown by Abrikosov, a linear band dispersion can lead to a linear MR in the quantum limit\cite{Abrikosov_1969,Abrikosov_1998,Abrikosov_2000}. In the case of \Ba122, the linear MR could be naturally explained by the presence of Dirac pockets in the AFM reconstructed state due to the symmetry protected band crossing\cite{Ran_2009}. Since the Dirac pockets have a small volume and a long mean free path\cite{Analytis_2009,Morinari_2010}, the high field limit can be reached with moderate fields, which could account for the non-linear Hall effect observed for the parent compound described above.

\begin{figure}
\includegraphics[width=8.5cm]{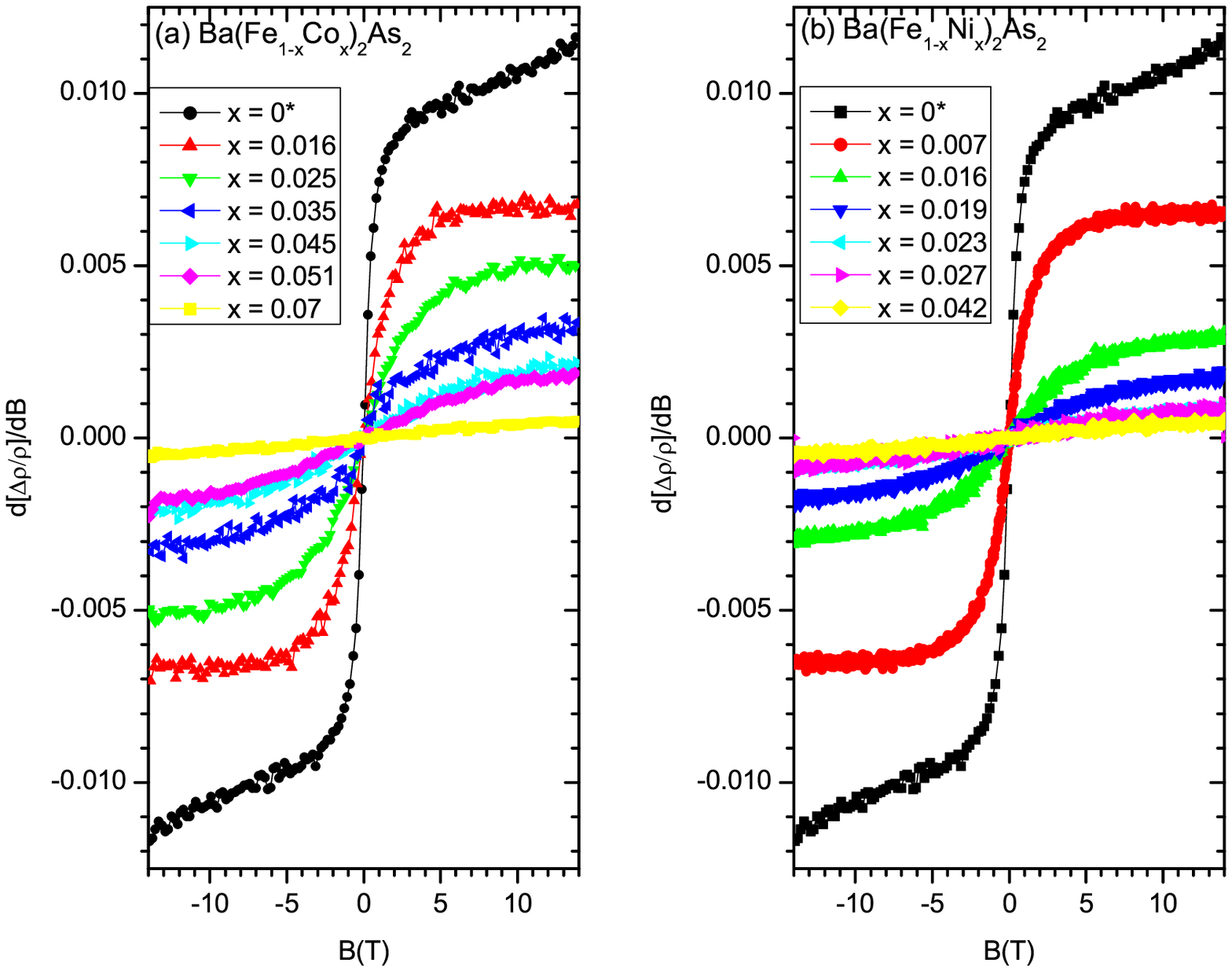}
\includegraphics[width=8.5cm]{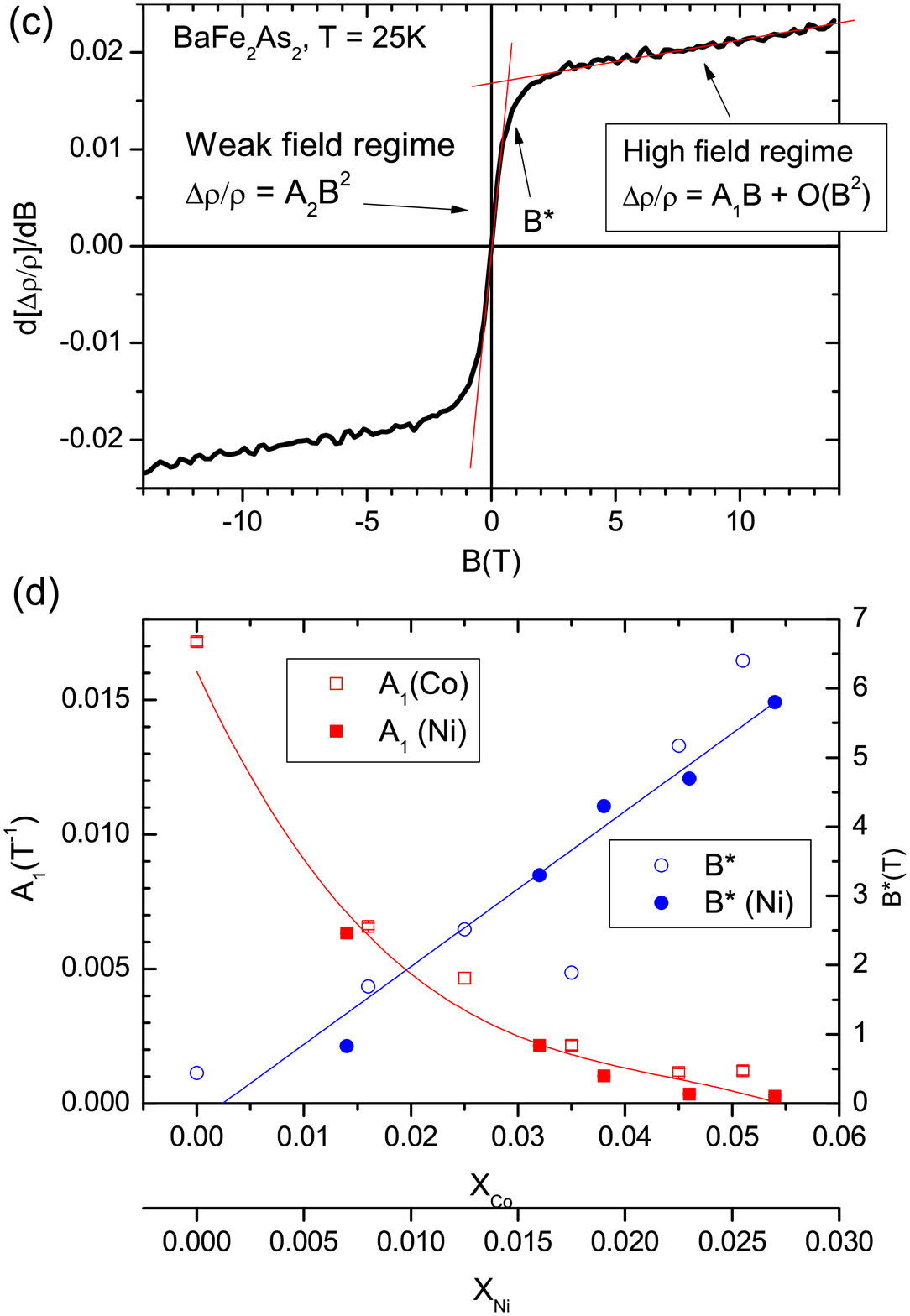}
\caption{\label{Fig:MRderivative} (Color online) (a) The field derivative of $MR$ of twinned samples of \CoBa122 and (b) \NiBa122 *Data for $x$ = 0 has been scaled down by a factor of two for clarity. (c) The field derivative of $MR$ ($dMR/dB$) measured on a twinned sample of the parent compound at $T = 25K$. A critical field scale $B^*$ clearly divides the $MR$ behavior into two regimes: below $B^*$ the $MR$ shows a weak field quadratic behavior and above $B^*$ $MR$ shows a high field linear behavior. (d) The doping evolution of the field scale $B^*$ (blue circles) and the high field $MR$ linear coefficient $A_1$ (red squares), for \CoBa122 and \NiBa122 (solid and open symbols respectively). }
\end{figure}

The cross-over from the weak-field $B^2$ dependence to the high-field linear dependence can be best seen by considering the field derivative of the MR, ($d[\Delta\rho/\rho]/dB$), which is plotted in Fig.\ref{Fig:MRderivative} for the parent compound. At low fields, $\Delta\rho/\rho = A_2B^2$, resulting in a linear field dependence for  $d[\Delta\rho/\rho]/dB$ as $B$ approaches zero. However, above a characteristic field $B^*$, $d[\Delta\rho/\rho]/dB$ starts to deviate from this weak field behavior, and appears to saturate to a much reduced slope. This indicates that at high field the MR is dominated by a linear field dependence, but there is also a small quadratic term ($\Delta\rho/\rho = A_1B + O(B^2)$).

Substitution of Co or Ni rapidly suppresses the linear $MR$ observed for the parent compound. Representative data are shown in Fig.\ref{Fig:MR}(b) for a detwinned single crystal of \CoBa122 with $x$ = 0.035 at 25 K. As can be seen, a linear MR is still observed, but the weak-field quadratic behavior extends to a higher field values. As for the parent compound, the anisotropy in the linear slope can be mainly accounted for by the anisotropy in $\rho_a$ and $\rho_b$. 

A comprehensive doping-dependence was obtained for twinned samples of \CoBa122 and \NiBa122. Representative data are shown in Figs. \ref{Fig:MR} (d) and (e) respectively, illustrating the rapid suppression of the MR with substitution. As described for the parent compound, we can extract the characteristic field $B^*$ by considering the field derivative of the $\Delta\rho/\rho$ (Fig. \ref{Fig:MRderivative}). Fitting the high field ($B > B^*$) MR by a second order polynomial, we obtain the coefficient of the linear field dependence coefficient $A_1$, the doping dependence of which is shown in Fig. \ref{Fig:MRderivative}. As can be seen, the characteristic field scale $B^*$ increases rapidly as a function of doping, whereas the linear coefficient decreases and almost vanishes at $x = 0.051$ for \CoBa122 and $x = 0.027$ for \CoBa122. Apparently, the rate at which the linear MR is suppressed is twice as rapid for Ni substitution as for Co substitution. This behavior is clearly correlated with the suppression of non-linearity in Hall effect, which occurs over a similar range of compositions as shown in Fig. \ref{Rxy}, providing additional evidence that the non-linear Hall coefficient in the parent compound is due to the presence of high-mobility Dirac pocket(s). 

\section{Discussion}

\begin{figure}
\includegraphics[width=8.5cm]{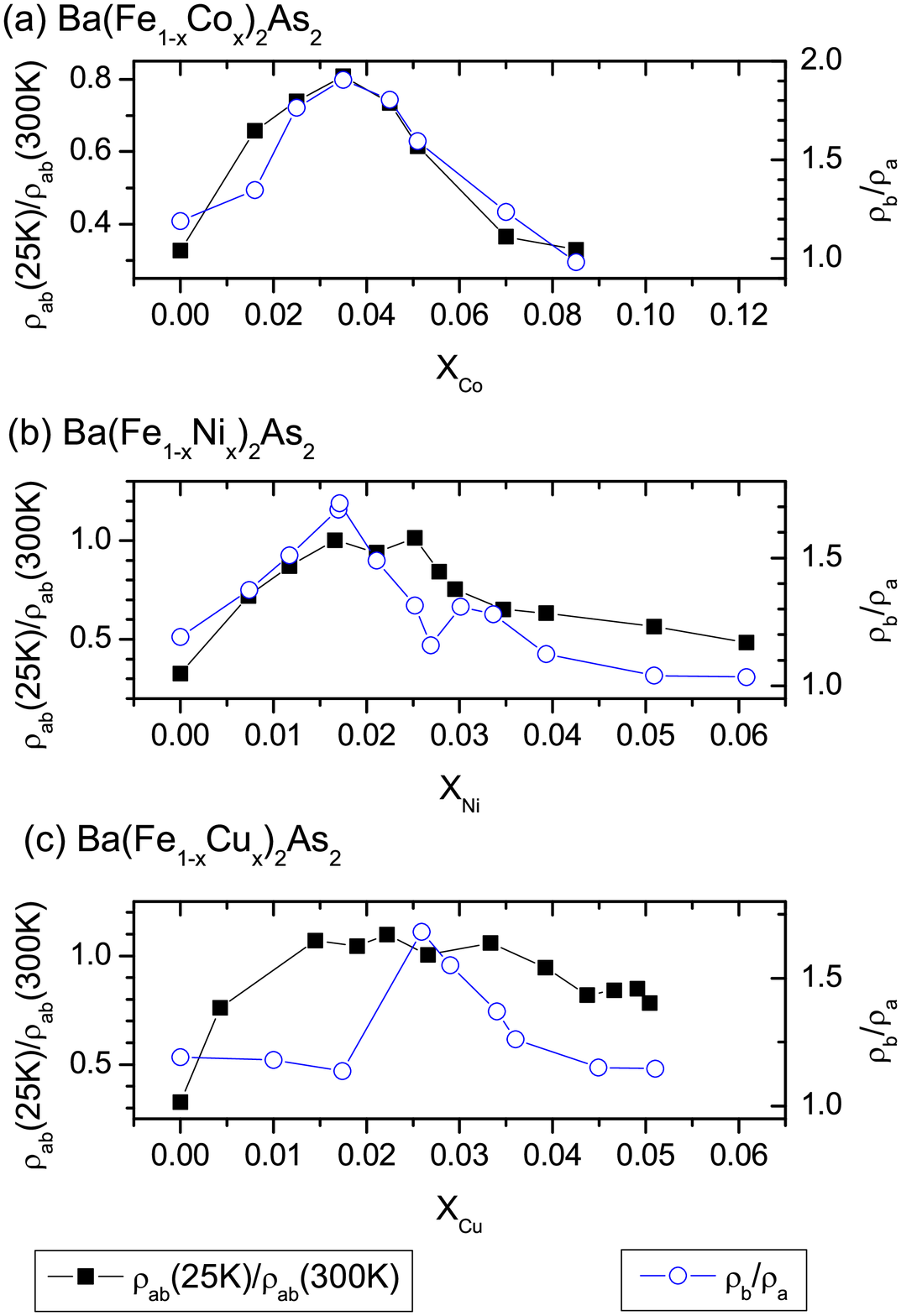}
\caption{\label{Fig:rho0} (Color online) The doping evolution of the in-plane resistivity of the twinned crystals at $T = 25K$ normalized by its room temperature value and the in-plane resistivity anisotropy ratio $\rho_b/\rho_a$ of detwinned crystal at $T = 25K$. Data are plotted for (a)\CoBa122, (b)\NiBa122, and (c)\CuBa122. }
\end{figure}

All three substitutions investigated in this study exhibit a non-monotonic variation of the in-plane resistivity anisotropy as the dopant concentration is progressively increased (Fig. \ref{CoNiCu}). Without further information it is not clear whether this effect is related to changes in the anisotropy of the scattering rate, or to changes in the electronic structure. However, as described above, consideration of the magnetotransport properties is suggestive of an important role for the Dirac pocket of the reconstructed Fermi surface. The progressive suppression of the linear MR with chemical substitution, and the associated suppression of the non-linear Hall coefficient, point to a scenario in which the contribution to the conductivity from the FS pockets associated with the protected band crossing (the Dirac pockets) rapidly diminishes with increasing dopant concentrations. For a multi-band system, the conductivity tensor is the sum of the contribution from each Fermi surface. If one particular Fermi surface pocket dominates the conductivity tensor, then the transport anisotropy will also be determined by the anisotropy of that particular Fermi surface. As observed previously by photoemission measurements, the Dirac pockets have an almost isotropic in-plane Fermi velocity \cite{Richard_2010}. If the mobility of this pocket is such that it dominates the transport, it would severely diminish the anisotropy associated with any other Fermi surfaces, just as we observe for the parent compound. The contribution from these Dirac pockets is progressively weakened by the transition metal substitution, which is manifested in the magnetoresistance and Hall effect. The subsequent emergence of a large in-plane resistivity anisotropy clearly indicates that the remaining low-mobility FS pockets are highly anisotropic. This change is also manifested in the normalized value of the residual resistivity at low temperature, which shows an abrupt increase at the same composition (Fig. \ref{Fig:rho0}).

The mechanism that suppresses the contribution from the Dirac pockets is unclear. The band crossing is protected by crystal inversion symmetry, but introducing impurities into FeAs planes locally breaks this symmetry. This effect would not only open a gap at the Dirac point, but would also increase the scattering rate for the Dirac electrons due to mixing of the orbital wavefunctions. This is consistent with the reduced mobility derived from the $MR$ analysis, and also consistent with the suppressed non-linearity in the Hall effect. On the other hand, Co substitution is argued to effectively electron-dope the system,\cite{Nakamura_2010} which shifts the chemical potential. If there is a gap at the Dirac point, this chemical potential shift could possibly lead to a Lifshitz transition in which the Dirac pocket vanishes. This possibility has been extensively discussed in the recent papers by Liu et al.\cite{Liu_2010,Liu_2011} A recent Nernst effect measurement also shows a suppression of Dirac transport by Co substitution in the \CoEu122 system\cite{Matusiak_2011}, suggesting the effect is not restricted to \Ba122 system.

To make a more quantitative understanding of how the Dirac pockets are being suppressed as a function of doping, one would ideally like to obtain the transport parameters for each band. According to Local Density Approximation (LDA) calculation, the \Ba122 parent compound has four closed Fermi surface pockets in the AF reconstructed states, therefore one would need eight independent parameters (mobility and carrier density for each pocket) to characterize the transport properties. If we focus only on the data for fields close to zero, we can obtain the coefficient of the $B^2$ quadratic term, $A_2$, by inverting the sum of the conductivity tensors of the four Fermi surfaces. We denote the conductivity and mobility of each Fermi surface by $\sigma_{i,j}$ and $\mu_{i,j}$, where the index $i = e,h$  stands for electrons or holes, and the index $j = D,P$ represents the Dirac bands and the parabolic bands.. We assume no intrinsic magnetoresistance for each individual Fermi surface, because in the condition of isotropic scattering rate and an ellipsoidal FS the leading quadratic term in Zener-Jones expansion is zero. The parameter $A_2$ is given by  
\begin{equation}
A_2 = \frac{\sigma_e\sigma_h(\mu_e+\mu_h)^2}{(\sigma_e + \sigma_h)^2} + \frac{\sigma_e}{\sigma_e + \sigma_h}A_{2,e} + \frac{\sigma_h}{\sigma_e + \sigma_h}A_{2,h}
\end{equation}
\begin{equation}
A_{2,i} = \frac{\sigma_{i,P}\sigma_{i,D}(\mu_{i,D}-\mu_{i,P})^2}{(\sigma_{i,P} + \sigma_{i,D})^2}
\end{equation}
\begin{equation}
\mu_i = \frac{\sigma_{i,D}}{\sigma_{i,P} + \sigma_{i,D}}\mu_{i,D} + \frac{\sigma_{i,P}}{\sigma_{i,P} + \sigma_{i,D}}\mu_{i,P}
\end{equation}
\begin{equation}
\sigma_i = \sigma_{i,D} + \sigma_{i,P}.
\end{equation}
Here, $\sigma_e,\sigma_h,\mu_e,\mu_h$ are the effective electron and hole conductivity and mobility in zero field. If we assume the the Dirac bands are dominating the transport, i.e. $\sigma_{e,D},\sigma_{h,D}\gg\sigma_{e,P},\sigma_{h,P}$ and $\mu_{e,D},\mu_{h,D}\gg\mu_{e,P},\mu_{h,D}$, the above expression can be greatly simplified because $\sigma_i\sim\sigma_{i,D}$ and $\mu_i\sim\mu_{i,D}$. The second and third term in equation 1 can be assumed to be much smaller than the first term:
\begin{eqnarray}
\frac{\sigma_e}{\sigma_e + \sigma_h}A_{2,e} = \frac{\sigma_e\sigma_{e,P}\sigma_{e,D}(\mu_{e,P}-\mu_{e,D})^2}{(\sigma_e + \sigma_h)(\sigma_{e,D} + \sigma_{e,P})^2}
\\ = \frac{\sigma_{e,P}\sigma_{e,D}(\mu_{e,P}-\mu_{e,D})^2}{(\sigma_e + \sigma_h)\sigma_e} \sim \frac{\sigma_{e,P}\mu_e^2}{(\sigma_e + \sigma_h)}\\
=\frac{\sigma_{e,P}(\sigma_e+\sigma_h)\mu_e^2}{(\sigma_e + \sigma_h)^2} \ll \frac{\sigma_e\sigma_h(\mu_e+\mu_h)^2}{(\sigma_e + \sigma_h)^2},
\end{eqnarray}
and equation 1 reduces to only the first term, which only depends on the effective mobility and conductivity of electrons and holes. The simplicity of this expression allow us to make a physical interpretation of the coefficient $A_2$ in terms of an effective mobility,
\begin{eqnarray}
\sqrt{A_2} = \frac{\sqrt{\sigma_e\sigma_h}}{\sigma_e + \sigma_h}(\mu_e+\mu_h) = \mu_{MR} \leq\frac{1}{2}(\mu_e+\mu_h)\equiv\mu_{ave}.
\end{eqnarray}
The square root of the quadratic field coefficient $A_2$, which we denote as $\mu_{MR}$, gives the lower bound of the average mobility of electrons and holes ($\mu_{ave}$). The equality between $\mu_{ave}$ and $\mu_{MR}$ holds only when the electron and hole conductivities are equal. However, even in the case of a strongly asymmetric conduction scenario, the $\mu_{MR}$ still gives a good estimate of $\mu_{ave}$. For example if $\sigma_e/\sigma_h = 10$ then $\mu_{MR} = 0.6\mu_{ave}$. All of the above reasoning makes sense only when the assumption that the Dirac bands dominate the transport holds. To be consistent, we extract the real number of the parent compounds to see if this is really the case. The value of $\mu_{MR}$ of the parent compounds at 25 K is about 1130 cm$^2$/Vs, which is comparable to the mobility of the Dirac pockets extracted from quantum oscillations for crystals prepared under similar conditions ($\sim$ 1000 cm$^2$/Vs)\cite{Analytis_2009}. This is indeed consistent with the assumption of $\mu_e\sim\mu_{e,D}$. We can also obtain the effective carrier density $n_{MR}$ by using the relation $\sigma = ne\mu$. If $\sigma_e\sim\sigma_{e,D}$, then $n_{MR}$ should be similar to the carrier density of Dirac carriers rather than the total carrier density. In the parent compound the observed value of $\mu_{MR}$ = 1130 cm$^2$/Vs corresponds to a very low effective carrier concentration, $n_{MR}$ = 0.003 electron per Fe. This value is much lower than the total number of carriers that one would obtain from the LDA calculations and ARPES measurements, but is comparable to the size of Dirac pockets as observed from the quantum oscillation experiments\cite{Analytis_2009}. Therefore the assumption $\sigma_e\sim\sigma_{e,D}$ also holds.
 
The effective mobility extracted from MR is much higher than the mobility directly obtained from the Hall coefficient $\mu_{Hall} = \vert R\sigma\vert = 376{\rm cm}^2/Vs$. This is because the contribution of electron and hole mobility in the Hall coefficient cancel each other:

\begin{eqnarray}
R = \frac{\sigma_e^2R_e + \sigma_h^2R_h}{(\sigma_e + \sigma_h)^2} = \frac{-\sigma_e\mu_e + \sigma_h\mu_h}{(\sigma_e + \sigma_h)^2}\\
\mu_{Hall} = \left\vert\frac{-\sigma_e\mu_e + \sigma_h\mu_h}{\sigma_e + \sigma_h}\right\vert\label{Eq:muHall}
\end{eqnarray}

In fact, the much smaller value of Hall mobility than the MR mobility already implies that the contribution to the conductivity of electrons and holes are of the same scale. To understand this one can consider the opposite case, one where electrons dominate the conduction $\mu_e \gg \mu_h$ and $\sigma_e  \gg \sigma_h$:

\begin{eqnarray}
\mu_{Hall} = \left\vert\frac{-\sigma_e\mu_e + \sigma_h\mu_h}{\sigma_e + \sigma_h}\right\vert \sim \left\vert\frac{-\sigma_e\mu_e}{\sigma_e + \sigma_h}\right\vert\sim \mu_e\\
\mu_{MR} = \frac{\sqrt{\sigma_e\sigma_h}}{\sigma_e + \sigma_h}(\mu_e+\mu_h)\sim \frac{\sqrt{\sigma_e\sigma_h}}{\sigma_e}\mu_e = \sqrt{\frac{\sigma_h}{\sigma_e}}\mu_e\label{Eq:muMR}\\
\sqrt{\frac{\sigma_h}{\sigma_e}} \ll 1 \Rightarrow \mu_{MR} \ll \mu_{Hall}
\end{eqnarray}
Therefore our measured value suggests that both electrons and holes play an important role in the transport in the reconstructed state in the parent compound, and this also gives us the confidence that $\mu_{MR}$ is a good estimate of $\mu_{ave}$, since their difference is smaller when $\sigma_e$ gets closer to $\sigma_h$. 

\begin{figure}
\includegraphics[width=8.5cm]{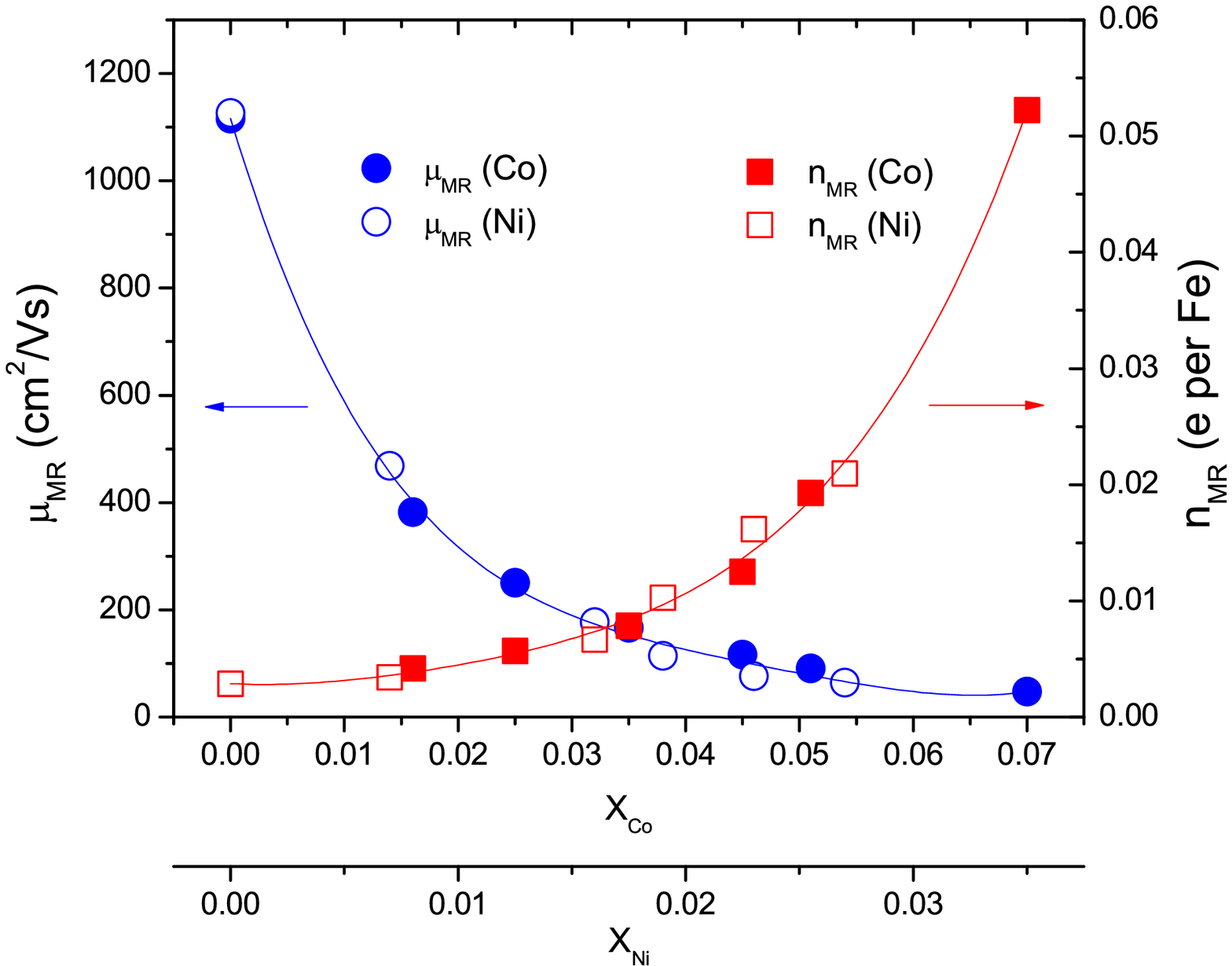}
\caption{\label{Fig:Mobility} (Color online) The doping evolution of the effective $MR$ mobility $\mu_{MR}$ and effective $MR$ carrier density $n_{MR}$ of \CoBa122 and \NiBa122 extracted from weak field $MR$, as described in the main text. }
\end{figure}

 As we argued above: the high $\mu_{MR}$ and small $n_{MR}$ reflects the fact that transport in the parent compound is dominated by a small number of high mobility carriers, i.e. the carriers from the Dirac pockets. However, this is no longer the case as we increase the Ni or Co doping concentration. The extracted value of $\mu_{MR}$ and $n_{MR}$ as a function of doping is plotted in Fig. \ref{Fig:Mobility}. By increasing doping concentration $\mu_{MR}$ decreases rapidly and $n_{MR}$ increases rapidly. Apparently our previous assumption that helped us simplify the $MR$ expression is no longer valid, and it is difficult to make a simple physical interpretation of the extracted $\mu_{MR}$ or $n_{MR}$. Nevertheless, the observed doping evolution is highly suggestive of a shift of the dominant role in transport to the high carrier density and low mobility carriers. Again the effect of Ni doping on suppressing the Dirac carriers is twice as fast as Co. Our data reveal that as the conductivity from the Dirac pockets is progressively suppressed, a large in-plane resistivity anisotropy emerges. The direct implication is that the other pockets of reconstructed FS are highly anisotropic, which is borne out by recent quantum oscillation measurements of \Ba122.\cite{Terashima_2011}  

Evidence for the suppression of the contribution of a high mobility pocket of reconstructed FS can also be found in the doping-dependence of the resistivity at low temperature. A direct comparison of the magnitude of the in-plane resistivity normalized by its room temperature value\footnote{To avoid uncertainty in geometric factor, the magnitude of the resistivity is normalized by its room temperature value, which is typically 250 $\sim$ 300 $\mu\Omega cm$ in this doping range\cite{Tanatar_2010b}. The room temperature value decreases slightly as a function of doping in the doping range we study here (less than 20\%), therefore the large variation of the normalized value at low temperature reflects the evolution of the electronic structure and charge dynamics.} \footnote{We used the resistivity of the twinned crystals, which is essentially the average of $\rho_a$ and $\rho_b$. The average value is actually more revealing to study the effect of isotropic Dirac pocket} and the in-plane resistivity anisotropy  at $T$ = 25 K is plotted if Fig. \ref{Fig:rho0}. As can be seen, the two quantities follow each other closely in the case of Co and Ni doping, but not for Cu substitution. Based on the previous analysis, the doping evolution of the resistivity can be readily understood. Initial suppression of the contribution to the transport arising from the Dirac pocket leads to an initial rise of the normalized resistivity. With progressive doping the magnetic order is further suppressed, releasing carriers and hence leading to a decrease in the resistivity for higher dopant concentrations. Note that the progressive doping also suppresses the structural transition that breaks the rotational symmetry, therefore after reaching a maximum value, the in-plane anisotropy also decreases.
  
We now comment briefly on the trends revealed by comparison of Co, Ni and Cu substitution. First principles density functional calculations indicate that the additional charge associated with Co and Ni substitution in \Ba122 resides within the muffin-tin potential \cite{Sawatzky_2010}, but an associated rigid-band shift leads to an effect that is ultimately equivalent to electron doping \cite{Nakamura_2010}. In calculations for transition metal impurities in LaFeAsO, Ni is found to be approximately twice as effective as Co in terms of the rigid band shift \cite{Nakamura_2010}, motivating a direct comparison of results for \CoBa122 and \NiBa122. In contrast, similar calculations for Zn and Cu impurities appear to strongly affect the electronic structure \cite{Nakamura_2010,Ryo_2011}, possibly accounting for significant differences between the phase diagrams of \CuBa122 and those of Co and Ni substituted \Ba122, including the much weaker superconductivity in Cu-substituted \Ba122 \cite{Ni_2010}. As anticipated, Ni substitution suppresses the contribution of Dirac carriers to the transport. Furthermore, this suppression occurs twice as rapidly with $x$ for Ni as for Co substitution. Similarly, the onset of the large in-plane resistivity anisotropy is found to occur for approximately half of the values of $x$ than is the case for Co substitution (Fig.\ref{CoNiCu}). The origin of the secondary maximum in the in-plane anisotropy observed for \NiBa122 is unclear, but perhaps reflects subtle changes in the reconstructed FS with chemical substitution. 

Inspection of Fig.\ref{Rxy} reveals that Cu substitution suppresses the non-linear Hall coefficient even more rapidly than Ni substitution. This effect is consistent with the deeper impurity potential of Cu relative to Ni, acting to increase the elastic scattering rate more rapidly per impurity added. However, the onset of the large in-plane resistivity anisotropy occurs for somewhat larger compositions, peaking at $x \sim 0.024$. Given that the effects of band filling and scattering are believed to be somewhat different for Cu substitution relative to Co or Ni substitution, it is perhaps not surprising to find that the transport properties evolve in a slightly different manner. Nevertheless, it is interesting to note that despite this, a region of the phase diagram still exists over which a large in-plane resistivity anisotropy is observed, presumably because high mobility isotropic FS pockets have been suppressed. Perhaps significantly, the compositional range over which this anisotropy is observed is narrower for Cu substitution than it is for Ni substitution, which in turn is somewhat narrower than for Co substitution. 

Finally, an intriguing correspondence can be made between the iron pnictides and underdoped cuprates.\cite{Chang_2011} Recent Nernst measurements reveal a large in-plane electronic anisotropy onsets at the pseudogap temperature\cite{Daou_2010}. Further Hall coefficient and quantum oscillation measurements suggested at even lower temperatures broken translational symmetry causes a reconstruction the Fermi surface and a high mobility isotropic electron pocket emerges\cite{Leboeuf_2011}. This high mobility electron pocket dominates the low temperature transport. As a result, not only does the Hall effect change sign from positive to negative, but the large electronic anisotropy which onsets at pseudogap temperature is also reduced. It has also been suggested that at a critical doping the system could undergo a Lifshitz transition at which the high mobility electron pocket disappears, which is also accompanied by an enhancement of in-plane resistivity anisotropy\cite{Leboeuf_2011}.

\section{Conclusions}
Co, Ni and Cu substituted \Ba122 are all found to exhibit a large in-plane resistivity anisotropy over a certain compositional range on the underdoped side of the phase diagram. The non-monotonic variation in the resistivity anisotropy as the dopant concentration is increased is especially striking given the uniform suppression of the lattice orthorhombicity. The non-linear Hall coefficient and linear $MR$ which are observed for the parent compound \cite{Zentile_2009,Huynh_2010}, and which are likely associated with the Dirac pockets of the reconstructed FS, are suppressed with increasing dopant concentrations. Intriguingly, for both Co and Ni substitution, for which a direct comparison is motivated based both on their respective phase diagrams and also on first principles LDA calculations \cite{Nakamura_2010}, the large in-plane resistivity anisotropy is found to emerge over the same range of compositions at which the non-linear Hall and linear $MR$ are progressively suppressed. Consideration of this evidence suggests that the isotropic, high-mobility Dirac pockets revealed by dHvA \cite{Analytis_2009,Harrison_2009,Terashima_2011} ARPES \cite{Richard_2010} and magnetotransport measurements \cite{Huynh_2010}, might effectively mask the intrinsic in-plane transport anisotropy associated with the other pockets of reconstructed FS. Within such a scenario, only when the contribution to the conductivity from the Dirac pockets is suppressed can the underlying anisotropy be revealed in the transport, perhaps accounting for the non-monotonic doping dependence.

Finally, we note that it remains to be seen to what extent the presence of a large resistivity anisotropy is a generic feature of the phase diagram of Fe-pnictide superconductors. Recent measurements of K-substituted \Ba122 indicate that the hole doped analog may not exhibit a large in-plane anisotropy \cite{Ying_2010}. This might reflect differences in the effect of electron vs hole doping on the reconstructed FS, or perhaps differences in the elastic scattering rate, since chemical substitution away from the Fe plane will presumably have a weaker effect.

\section{Acknowledgments}

The authors thank C.-C. Chen and S. A. Kivelson for helpful discussions. This work is supported by the DOE, Office of Basic Energy Sciences, under contract no. DE-AC02-76SF00515. Part of the magnetotransport experiment was performed at the National High Magnetic Field Laboratory, which is supported by NSF Cooperative Agreement No. DMR-0654118, by the State of Florida, and by the DOE.

\end{document}